\documentclass[12pt]{article}
\usepackage{graphicx,amsmath}
\usepackage{units}
\usepackage{fancybox}

\newlength{\dinwidth}
\newlength{\dinmargin}
\def\lapproxeq{\lower .7ex\hbox{$\;\stackrel{\textstyle                                                    
<}{\sim}\;$}}                                                    
\def\gapproxeq{\lower .7ex\hbox{$\;\stackrel{\textstyle                                                    
>}{\sim}\;$}}                                                    
\def\be{\begin{equation}}                                                    
\def\ee{\end{equation}}                                                    
\def\bea{\begin{eqnarray}}                                                    
\def\eea{\end{eqnarray}}

\def\qq{q\bar{q}}

\def\sh{\hat s}
\def\sh2{{\hat s}^2}

\def\e{\epsilon}
\begin{document}

\begin{flushright}                                                    
IPPP/13/32  \\
DCPT/13/64 \\                                                    
\today \\                                                    
\end{flushright} 

\vspace*{0.5cm}

\begin{center}
{\Large \bf Treatment of the infrared contribution:\\}
\vspace{0.5cm}
{\Large \bf NLO QED evolution as a pedagogic example}

\vspace*{1cm}
                                                   
E.G. de Oliveira$^{a,b}$, A.D. Martin$^a$ and M.G. Ryskin$^{a,c}$  \\                                                    
                                                   
\vspace*{0.5cm}                                                    
$^a$ Institute for Particle Physics Phenomenology, University of Durham, Durham, DH1 3LE \\                                                   
$^b$ Instituto de F\'{\i}sica, Universidade de S\~{a}o Paulo, C.P.
66318,05315-970 S\~{a}o Paulo, Brazil \\
$^c$ Petersburg Nuclear Physics Institute, NRC Kurchatov Institute, Gatchina, St.~Petersburg, 188300, Russia \\          
                                                    
\vspace*{1cm}

\begin{abstract}                                                    

We show that the conventional prescription used for DGLAP parton evolution at NLO has an inconsistent treatment of the contribution from the infrared (IR) region. We illustrate the problem by studying the simple example of QED evolution, treating the electron and photon as partons. The deficiency is not present in a physical approach which removes the IR divergency and allows calculation in the normal 4-dimensional space.

\end{abstract}                                                        
\vspace*{0.5cm}                                                    
                                                    
\end{center}

\section{Introduction}

It was mentioned long ago that the treatment of the infrared singularity in next-to-leading order (NLO) QCD evolution may be ambiguous. Depending on the approach, one gets a different non-vanishing non-singular result after the cancellation of the infrared singularities. Moreover, sometimes the non-singular result may even be gauge dependent; see, for example, \cite{BvH,BvH2}. Since the predictions of observables should be unambiguous, clearly some improvement of the formalism is necessary.

Recall that actually any well-defined physical quantity should have {\em no} infrared (IR) singularities.  An IR singularity in a particular diagram will therefore  \\
 --- {\it either} be cancelled by the  contribution of another diagram,\\--- {\it or} be avoided by the cancellation between real soft gluon/photon emission and the virtual loop correction (of the same order in $\alpha$-coupling) to the main hard process,\\--- {\it or} not occur due to a physical cutoff of the large-distance divergency caused by the kinematics corresponding to a particular
experiment. 

We list a few examples:
\begin{itemize}
\item
The well-known Bloch-Nordsieck~\cite{BN} cancellation between the real emission of a soft photon which cannot be observed experimentally and the loop
correction (the Sudakov form factor) to the main process without extra soft photon emission. After the cancellation, the physical infrared cutoff is fixed by the experimental resolution, that is by the energy of photons which {\em can} be observed in the particular experiment.
\item  
An analogous cancellation between the low-$k_t$ gluon emission 
($\propto 1/k^2_t$) in BFKL evolution and gluon reggeization\footnote{The next ($\alpha_s\ln s$) power term in the BFKL amplitude can be generated either by the emission of a new $s$-channel gluon or it may come from the $\alpha_s$ decomposition of the $t$-channel gluon Regge trajectory. In the latter case the IR logarithmic divergency is hidden in the expression for the gluon trajectory and corresponds to the integration over the $t$-channel gluon virtuality in the loop which describes gluon reggeization. At first sight, these two expressions seem to have nothing to  do with each other. However, by changing the variable and using an identity between two integrals, it was shown that the Regge trajectory contribution may be written in a form which provides an exact cancellation of the $1/k^2_t$ singularity caused by the new low-$k_t$ gluon emission~\cite{BFKL}. Thus, finally the BFKL kernel has no IR divergency.}.
\item
The cross section of high energy quark-quark scattering caused by the two-gluon exchange (the Low-Nussinov Pomeron \cite{LN}) has a divergency of the form $dk^2_t/k^4_t$. This is eliminated for real colourless hadron scattering by accounting for the interaction with quark spectators.
\item
The logarithmic divergency in the NLO photon-gluon fusion DIS coefficient function is cancelled by subtracting the contribution already generated at LO by the convolution of the LO 
gluon-to-quark ($P_{qg}$) splitting function and the LO coefficient function~\cite{OMR}.  The subtraction is necessary to avoid double counting.
\end{itemize}

Thus each time we face an IR singularity we must either find how it is cancelled or find the true, physically-motivated, infrared cutoff\footnote{Note that in the case of QCD the large-distance contribution will be anyway limited by confinement.}. The result cannot be ambiguous or depend on one or another {\em prescription}. The dimensional ($D=4+\epsilon$) regularization used to eliminate the divergency in the ultraviolet region should not affect the IR domain since after we trace all the IR cancellations and physically-motivated cutoffs there can be no place for an IR contribution coming from very large distances.
 
Here we study the role of the IR (large-distance) domain during parton evolution. As we have argued above, this should cause no problem; there should be no IR divergency and we should obtain a unique prediction for any well-defined observable. Unfortunately the conventional prescription used for DGLAP parton evolution at NLO contains a non-zero contribution of IR ($\epsilon/\epsilon$) origin which makes no physical sense. In Section \ref{sec:2} we show that this contribution comes from an incorrect treatment of the subtraction of the term already generated by the two steps of the LO evolution from the diagrams corresponding to the NLO splitting. Then in Section \ref{sec:alt} we discuss whether or not the difference between the conventional $\overline{\rm MS}$ and `physical' prescriptions may be considered as an `alternative factorization scheme'. In Section \ref{sec:4} we explain the problem with the `conventional' treatment in more detail, and show how it can be avoided by using a `physical' approach.  In Section \ref{sec:5} we perform the calculations for QED evolution. 
  For QED there is no confinement and all the quantities are well-defined in terms of the electron and photon (regarded as partons). However, the extension to QCD is straightforward.

\section{Treatment of the IR region in NLO DGLAP evolution \label{sec:2}}

\begin{figure} [htb]
\begin{center}
\includegraphics[height=10cm]{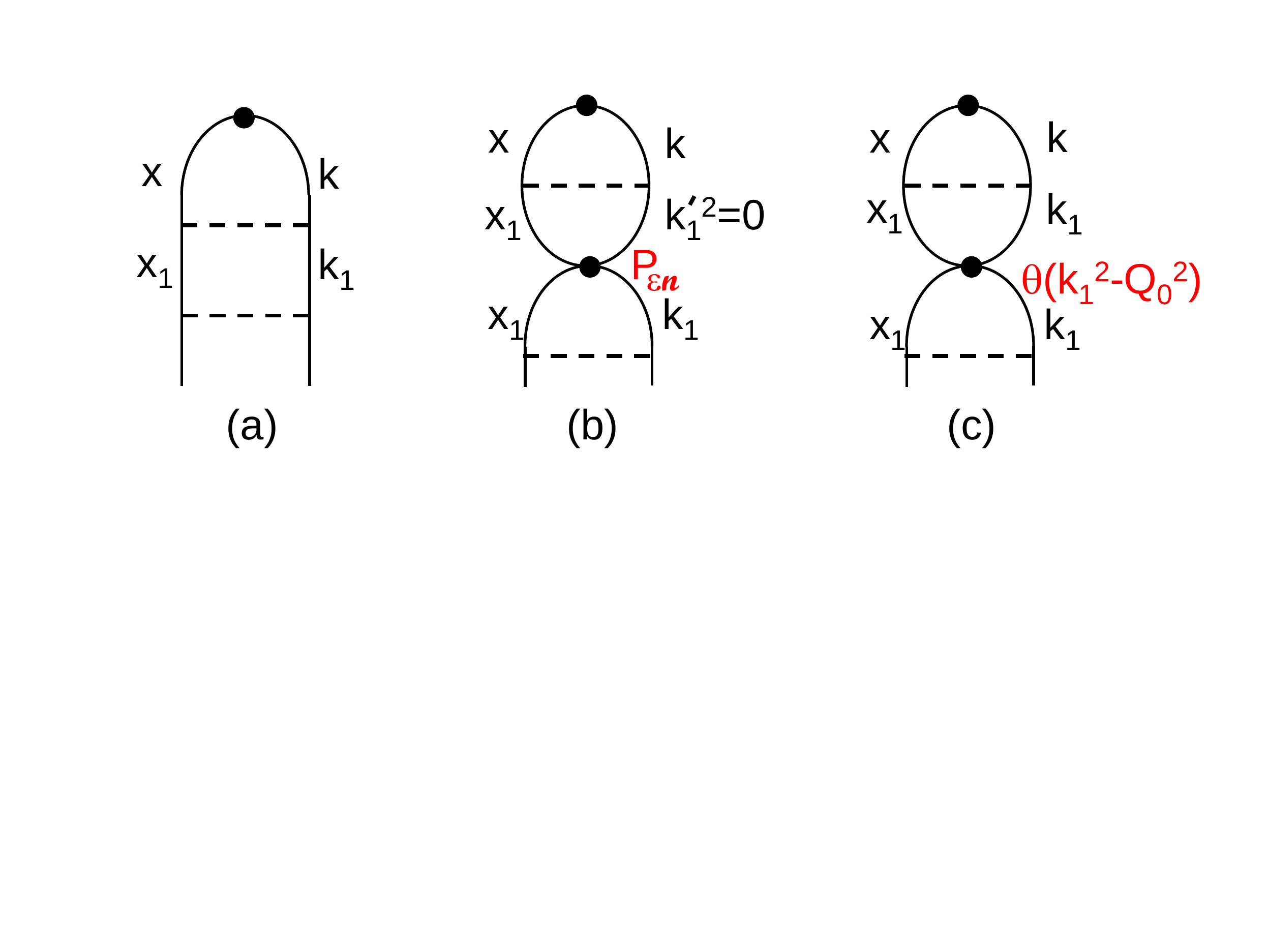}
\vspace*{-5.5cm}  
\caption{\sf (a) The ${\cal O}(\alpha^2)$ ladder diagram used to extract the NLO DGLAP splitting function; (b) the LO$\otimes$LO term that must be subtracted from (a) (in the conventional dimensional regularisation approach) to avoid double counting; (c) the LO$\otimes$LO diagram subtracted from (a) in the `physical' approach.  }
\label{fig:QEDfig1}
\end{center}
\end{figure}

  The problem appears for the first time at the NLO level. For LO DGLAP evolution we have strong $k_t$ ordering, and so starting from some non-zero virtuality $Q_0^2$ we never enter the infrared domain. It is convenient to use the axial gauge, since then the only infrared collinear divergency is logarithmic, and occurs only in the ladder (box) diagram\footnote{The divergency caused by soft photon (gluon) emission is cancelled by the usual Bloch-Nordsieck procedure encoded in the $1/(1-z)_+$ prescription.}.  At NLO the relevant diagram is shown in Fig. \ref{fig:QEDfig1}(a). This two-loop ladder diagram generates the  ${\cal O}(\alpha^2)$ contribution\footnote{We use the appropriate coupling $\alpha$  depending on whether we are studying QED or QCD.}  which consists of both the NLO part and a part generated by two-step LO evolution. Therefore to extract the NLO result we must subtract this LO$\otimes$LO contribution. Here we are concerned with two different approaches to performing the subtraction.
\vspace{-0.7cm}

\subsection{Conventional treatment}

First, we consider the conventional dimensional regularisation approach, with the corresponding LO$\otimes$LO subtraction shown by diagram (b). We follow
 the original paper for (non-singlet) NLO evolution by Curci, Furmanski and Petronzio \cite{CFP}.  The central dot in the LO$\otimes$LO diagram (b) represents a projector
 \be 
 P_{\e n}\equiv P_\epsilon\otimes P_n,
 \label{eq:Pen}
 \ee
  in momentum and spinor space, which ensures that both evolution steps in the diagram are at LO.  The projector puts the intermediate parton on-mass-shell ($k_1^{\prime 2}=0$) and keeps only the $1/\epsilon$ singular part of the lower cell, and only that spin structure which will produce the $1/\epsilon$ pole in the next cell.  
 However, there is a danger that some of the $\epsilon/\epsilon$ finite terms of IR origin which are present in diagram (a), are killed by the projector acting in diagram (b).  The crucial question is how to treat these $\epsilon/\epsilon$ terms.
  
Usually one does not need to worry about the IR  $\epsilon/\epsilon$ contribution coming from an individual diagram, since it is known that finally there is no IR divergency and so it is natural to expect that all the large-distance contributions will be cancelled. However, to provide an exact cancellation all terms should be calculated under the same conditions. For example if at LO level we have some additional boundary condition like $k_1^2>Q_0^2$, then the same condition should be applied to the NLO diagram.

\subsection{`Physical' treatment}

Recall that the only IR divergency at NLO level comes from the box (and self-energy) diagrams (in the axial gauge used to provide the usual partonic interpretation). The divergency is identical to that accounted for in LO evolution. The divergency has a logarithmic $dk_1^2/k_1^2$ form and comes from the region $k_1^2 \ll k^2$. Therefore, in this domain the contribution of the two-loop ladder diagram is exactly equal, up to non-singular ${\cal O}(k^2_1/k^2)$ corrections,  to that generated by the two-step LO evolution, Fig. \ref{fig:QEDfig1}(c). Thus, after the subtraction of this LO$\otimes$LO term, the remaining NLO part (that is the NLO splitting function) should not contain any large-distance contributions.  So a sensible {\it physical} way of proceeding, is to subtract the LO-generated contribution (diagram (c)) from the ladder-like Feynman graph (diagram (a)) before the integration. That is, work directly in terms of diagrams. Since, for $k^2 \ll Q^2$, the contribution of the NLO graph is exactly equal (up to ${\cal O}(k^2/Q^2)$ terms) to that generated by the LO diagrams, the infrared singularity is cancelled; and so the remaining part may be integrated in normal 4-dimensional space.
  
Note that, in LO evolution, we evolve upwards starting from some fixed virtuality $k^2_1 \ge Q^2_0$,  and due to strong $k_t$-ordering we never enter the region below $Q^2_0$. On the other hand, the original two-loop diagram samples this region. So we have a potential inconsistency.  If all low scale ($<Q_0$) effects are represented by the phenomenological starting distributions for the evolution, we have {\it either} to eliminate the contribution of low $k_1^2$ from the two-loop ladder diagram (that is, completely exclude the IR contribution), {\it or} to subtract the LO-generated part in its original exact form (Fig. \ref{fig:QEDfig1}(c)) without any additional limits or distortions caused by the projector. We will follow the second approach. As was argued above, such a subtraction completely eliminates the IR contribution from the two-loop diagram, leading to a final result which does not contain finite $\epsilon/\epsilon$ terms of IR origin. Since after the  LO$\otimes$LO subtraction, the NLO part has no IR divergency\footnote{The divergency is completely removed, leaving only power-suppressed ${\cal O}(Q^2_0/Q^2)$ terms.}, the result can be calculated directly in normal $D=4$ dimensional space. We call this the `physical' approach.

\subsection{Discussion}  
Unlike the `physical' approach, in `conventional'  dimensional regularisation we work in $4+\epsilon$ dimensions. Here, the LO$\otimes$LO subtraction generated by the LO evolution is distorted by the projection operator, $P_{\epsilon n}$, see Fig. \ref{fig:QEDfig1}(b). This projector simplifies the calculation of the leading $1/\epsilon$ pole contributions, but does not provide an exact cancellation of the IR part, leaving some unphysical finite  $\epsilon/\epsilon$  terms.
 However, to provide an exact cancellation of the unphysical large-distance contributions we must subtract these non-singular terms as well.

We have checked that if we would have kept the correct $\epsilon/\epsilon$ terms in the LO$\otimes$LO subtraction in the dimensional regularisation approach, then we would have reproduced the result obtained in the `physical' 4-dimensional approach, for the NLO coefficient functions for deep-inelastic and Drell-Yan processes \cite{OMR}. Below we denote this discrepancy in the NLO coefficient functions by $\Delta C$.

In summary, the subtraction based on Feynman diagrams completely cancels the infrared divergency, so there is no need to use infrared regularisation.  As a consequence, there are no sources of infrared ambiguity. We emphasize that this `physical' prescription differs from the `conventional' ${\overline {\rm MS}}$ approach by the non-neglect of the contribution of the $\epsilon/\epsilon$ terms.

\section{Can the difference be regarded as a scheme change?   \label{sec:alt}}
Let us start with discrepancy $\Delta C$ which originates from the $\e/\e$ infrared contribution to the non-singlet coefficient function, $C^{\rm NLO}$, in QED. We have checked, by explicit calculation, 
 that the function 
\begin{equation}
\label{eq:dc}
\Delta C~=~\frac{1+x^2}{1-x} \ln (1-x) + (1-x) - \frac{9}{4} \delta(1-x),
\end{equation}
where the `+' prescription is implied.
It originates from the $\cal O(\e)$ terms in the LO splitting function
\be
P(x,\e)=P^{\rm LO}(x)+\delta_P(x)\ ,
\ee
where  $\delta_P$ 
is the part of the LO splitting function, proportional to $\e$.
 Explicitly, it is given by~\cite{CFP}
\begin{equation}
\delta_P(x)~=~\epsilon (1-x)+\epsilon P^{\rm LO}(x){\rm ln}(1-x)-\epsilon\frac{9}{4} \delta(1-x).
\label{eq:3terms}
\end{equation} 
The first term comes from the extra photon polarisation states in $D=4+\epsilon$ space. The second term arises from the phase space factor $(k^2_{t1})^\epsilon$, which, when expressed in terms of virtuality variables, takes the form
\begin{equation}
(k^2_{t1})^\epsilon~=~(k^2_{1}(1-x))^\epsilon ~=~1+\epsilon{\rm ln}k^2_1+\epsilon {\rm ln}(1-x).
\end{equation}
The third term in (\ref{eq:3terms}), involving $9/4$, arises from the self-energy diagram. It is necessary to keep this term to satisfy the charge/flavour   conservation law $\int\delta_P (x) dx=0$.

Thus expression (\ref{eq:dc}), which represents the difference between `conventional' and `physical' treatments of the IR region for the (non-singlet) NLO coefficient function, is just the convolution of the LO
 coefficient function, $C^{\rm LO}=\delta(1-x)$, with the part, $\delta_P$, of the LO splitting function that is proportional to $\e$.

We note that the correction to the NLO coefficient function, $\Delta C$, may be absorbed by redefining the parton distribution
\begin{equation}
e'(z)~=~e(z)+\frac{\alpha}{2\pi}\int\frac{dx}{x}\Delta C(x)~e(z/x)~\equiv~e+  \frac{\alpha}{2\pi}\Delta C\otimes e.
\label{eq:eprime}
\end{equation}
That is, at first sight the $\Delta C$ effect may be considered as an `alternative factorization scheme'. Then the evolution equation for $e'(z)$ takes the form
\begin{eqnarray}
\frac{de'(z,Q^2)}{d{\rm ln}Q^2}&~=~& \frac{\alpha}{2\pi} P \otimes e+ \left(\frac{\alpha}{2\pi}\right)^2 \Delta C\otimes P \otimes e  \\
  &~=~&  \frac{\alpha}{2\pi}P\otimes e'+ \left(\frac{\alpha}{2\pi}\right)^2[\Delta C,P]\otimes e'+{\cal O}(\alpha^3),
\end{eqnarray}
where here $P=P^{\rm LO}+P^{\rm NLO}$. 
Thus provided that the analogous 
contribution to the splitting function $\Delta P(x)$ is equal to commutator 
\begin{equation}
[\Delta C,P]~\equiv~\Delta C\otimes P-P\otimes \Delta C~
\end{equation}
we will have the correct non-singlet NLO evolution.  

However, since this commutator vanishes we must have $\Delta P=0$ for non-singlet evolution in the new factorization scheme. If, on  the contrary, we find a non-zero difference $\Delta P$ between the splitting functions calculated within the `physical' and the `conventional' approaches, it would mean that
the correct `physical' treatment of the IR domain {\it cannot} be considered as the `conventional' treatment in an alternative factorization scheme.  So, clearly,  we have to study the difference $\Delta P$.

\section{The NLO splitting function in QED \label{sec:4}}
 
 To investigate $\Delta P$ we consider non-singlet NLO evolution in well-defined QED theory; that is we compare the cross section for $\gamma^* e \to eX$ calculated using first the `conventional' treatment and then the `physical' treatment of the IR region.  For such evaluations we need to calculate both the NLO splitting function and the NLO non-singlet coefficient function.\footnote{For the $\gamma^*g \to \qq$ DIS coefficient function and the $qg \to q\gamma^*$ Drell-Yan coefficient function of QCD, we have explicitly checked \cite{OMR} that the $\epsilon/\epsilon$-corrected conventional result exactly reproduces the result  obtained in the `physical' approach working in normal 4-dimensional space. Recall that in the physical approach the ladder diagram generated by the LO splitting function is subtracted from the corresponding Feynman diagram with the result that the infrared singularity is exactly cancelled.}

Since we evolve upwards from a starting scale $Q^2_0 \gg m^2_e$, we can therefore neglect $m_e$. For the conventional approach we refer to the original well-known work of Curci, Furmanski and Petronzio \cite{CFP}.\footnote{Although \cite{CFP} was written for QCD, it applies equally well to QED. The only simplification is that there is no equivalent of the triple-gluon vertex, and so only the $C_F^2$ (and $N_F C_F$) diagrams in Table 1 of \cite{CFP} are relevant for QED.} The only place where the $\epsilon/\epsilon$  infrared problem occurs is in the subtraction of the LO-generated contribution shown by the two diagrams at the head of the fourth column of Table 1 of \cite{CFP} (or equivalently, diagrams (a)-(b) of Fig. \ref{fig:QEDfig1}). All other $C_F^2$ diagrams in Table 1 have no collinear infrared divergency; the soft photon divergencies, encoded in $I_1$ and $I_0$, are cancelled between pairs of diagrams (see the last four entries in Table 1 of \cite{CFP}), as is expected from the Bloch-Nordsieck theorem.

\subsection{The LO$\otimes$LO subtraction}
We emphasize again that to completely remove the contribution coming from very large distances in Fig. \ref{fig:QEDfig1}(a), we must keep the $\epsilon/\epsilon$ terms generated by the LO splitting function, that is by diagram (b). Note that in the original LO evolution we start from a non-zero scale $Q_0$. There, the ${\cal O}(\epsilon)$ contribution to the splitting function is negligible, since we have an explicit cutoff, $Q^2>Q^2_0$, and never reach the IR $1/\epsilon$ pole.
The problem first occurs in the ${\cal O}(\alpha^2)$ diagram (a) of  Fig. \ref{fig:QEDfig1}; the ladder integral has no such cutoff; it runs down to $k_1^2=0$, and so generates a $1/\epsilon$ pole. This pole is cancelled by subtracting the LO$\otimes$LO term generated by normal LO evolution, diagram (b). However the constant $\epsilon/\epsilon$ terms were not cancelled in \cite{CFP} completely. To be consistent, we should {\it either} include the cutoff $|k^2|>Q_0^2$, analogous to that in LO evolution, {\it or} continue the integral generated by the LO splitting function down to $|k^2|=0$.  In both cases we would completely eliminate the infrared contribution.

To  extract the NLO splitting function we must therefore subtract from the ${\cal O}(\alpha^2)$ contribution (arising from the ladder diagram of Fig. \ref{fig:QEDfig1}(a)) the part that is already generated by the LO$\otimes$LO convolution shown in diagram (b) in the conventional approach, or shown in diagram (c) in the `physical' treatment. In either case the result may be written in the form
\be
\int^{\mu^2} \frac{dk^2}{k^2} \int^{k^2 x_1/x} dk^2_1\int \frac{dx_1}{x_1}\left[\frac{A}{k^2_1} +\frac{B}{k^2}\right] ~-~ \int^{\mu^2} \frac{dk^2}{k^2} \int^{k^2} dk^2_1\int \frac{dx_1}{x_1}~\left[\frac{A^{\prime}}{k^2_1}\right]
\label{eq:a1}
\ee 
 where $A,~A^{\prime}$ and $B$ are known functions of $x$ and $x_1$. In particular\footnote{For completeness, we have $B=-x/x_1-x_1x +x^2/x_1^2-x^2/x_1-2x^3/x_1^3+x^3/x_1^2-x^3/x_1.$}
\be
A'=P^{\rm LO}(x/x_1)~ P^{\rm LO}(x_1).
\label{eq:a2}
\ee

\subsection{The role of the $P_{\epsilon n}$ projector}
First, we consider the subtraction in the conventional dimensional regularisation approach working in $D=4+\e$ space; that is the subtraction of diagram \ref{fig:QEDfig1}(b), After the integrations over $k^2$ and $k^2_1$ the $A$ and $A^{\prime}$ terms in (\ref{eq:a1}) produce double and single pole contributions, $1/\e^2$ and $1/\e$ of infrared origin, while $B$ gives just a single pole. According to the dimensional regularisation prescription the leading double pole is crossed out. Our interest is the subtraction of the LO$\otimes$LO $1/\e$ term generated by 
$ A^{\prime}$, from the corresponding term in $ A$ arising from the ${\cal O}(\alpha^2)$ ladder diagram. The form of the $A^{\prime}$ term is the same as $A$ of (\ref{eq:a2}), except for the projector $P_{\e n}$ of (\ref{eq:Pen}) acting on the lower cell of Fig. \ref{fig:QEDfig1}(b), which gives different input for the upper cell, that is it puts the `upper' electron on-shell, $k_1^2 \to k_1^{\prime 2}=0$, and reduces the $dk^2_1$ integration to simply $1/\e$. At first sight, this prescription, seems to guarantee the cancellation of the single-pole contributions  generated by the large-distance (infrared) domain of $k^2_t\to 0$.

Indeed,
it is the single pole contribution of the form $\e/\e^2$, coming from the part of the LO splitting function, $\delta_P $ proportional to $\e$, which is of interest here, so we
 study this single pole contribution  in more detail. For the $A$ 
 term the $dk^2$ and $dk^2_1$ integrations of the ladder graph of 
Fig.\ref{fig:QEDfig1}(a) give
\be
\int^{\mu^2}\frac{dk^2}{(k^2)^{1-\e}}~\int^{x_1k^2/x}\frac{dk_1^2}{(k^2_1)^{1-\e}} ~=~ \left(\frac{x_1}{x}\right)^{\e} \frac{(\mu^2)^{2\e}}{2\e^2},
\ee
 whereas the presence of the projector $P_{\e n}$ in the $A^{\prime}$ term 
 effectively gives a factor of 2 enhancement~\footnote{Here we omit the $\e$ dependence coming from the factor $(x_1/x)^\e\simeq 1+\e\ln(x_1/x)\simeq 1$. The last term, $\e\ln(x_1/x)$, comes from the interval of a rather large $k^2_1$, from $k^2$ up to $k^2x_1/x$. It is a normal NLO contribution which has nothing to do with the IR domain.}
 \be
\int^{\mu^2}\frac{dk^2}{(k^2)^{1-\e}}~P_{\e n}\int^{x_1k^2/x}\frac{dk_1^2}{(k^2_1)^{1-\e}} ~=~ \left(\int^{\mu^2}\frac{dk^2}{(k^2)^{1-\e}}\right) \frac{1}{\e}~=~ \frac{(\mu^2)^{\e}}{\e^2}.
\label{eq:a3}
\ee

Now consider the contribution, $\Delta P$, to the NLO splitting function that is proportional to $\epsilon$ arising from the IR region, $k_1^2 \to 0$. This comes from convolutions of the additional part, $\delta_P$, of the LO splitting function calculated in $D=4+\e$ space at one step of the LO evolution, with the LO splitting function of the other step of the evolution. That is, from diagrams (a) and (b) of Fig. \ref{fig:QEDfig1} we have
\begin{equation}
\Delta P~=~\left[\delta_P (x_1)\otimes P^{\rm LO}(x/x_1)+P^{\rm LO}(x_1)\otimes \delta_P (x/x_1)\right]~~-~~\left[0~+~2P^{\rm LO}(x_1)\otimes \delta_P(x/x_1)\right].
\label{eq:a4}
\end{equation}
Here the first [....] is the $\e$ contribution of the $A$ term in 
(\ref{eq:a1}). The second [....] is from the LO$\otimes$LO subtraction term $A^{\prime}$ in (\ref{eq:a1}), where the projector, $P_{\e n}$, kills the contribution from the lower cell, and doubles the contribution from the upper cell, see (\ref{eq:a3}). 
Clearly the projector $P_{\e n}$ leads to physical differences between the `conventional dimensional regularisation' and the `physical' treatments of the IR region.

Let us return to the $\e/\e^2$ contribution of (\ref{eq:a4}), which has the general form
\begin{equation}
\Delta P~=~\left[\delta_P (x_1)\otimes P^{\rm LO}(x/x_1)~-~P^{\rm LO}(x_1)\otimes \delta_P (x/x_1)\right].
\label{eq:a5}
\end{equation}
If $\Delta P \ne 0$ then there would be an $\e/\e^2$ term missing in the conventional approach, whereas if $\Delta P=0$ then we could reach agreement with the `physical' treatment of the IR region by a change of factorisation scheme. Is $\Delta P$ of (\ref{eq:a5}) zero or non-zero? At first sight, it looks as if $\Delta P=0$ due to the $x/x_1 \Leftrightarrow x_1$ symmetry of the $dx_1/x_1$ integration. If this symmetry were to hold, then the $dx_1$ integrations over the two terms in (\ref{eq:a5}) would be equal and $\Delta P$ would vanish.

However, the symmetry is violated by the different forms of the ``+'' prescription\footnote{ By the ``+'' prescription we mean
$\int^1_0 dzf(z)/(1-z)_+=\int^1_0 dz(f(z)-f(1))/(1-z),$ with $1/(1-z)_+=1/(1-z)$ for $0 \le z<1$.}
used in the second step of the evolution; a difference which generates a non-physical $\e/\e$ contribution, $\Delta P$, to the NLO splitting function. The LO$\otimes$LO term in (\ref{eq:a1}) has the general structure
\be
\int\frac{dx_1}{x_1} ~\left(\frac{1+x_1^2}{(1-x_1)_+}+\e \delta_P(x_1)\right)\left(\frac{1+(x/x_1)^2}{(1-x/x_1)_+}+\e~ {\rm terms}\right),
\label{eq:a7}
\ee
whereas for the relevant part of first term in (\ref{eq:a1}) we have
\be
\int dx_1 ~\left(\frac{1+x_1^2}{(1-x_1)_+}+\e \delta_P(x_1)\right)\left(\frac{1+(x/x_1)^2}{(x_1-x)_+}+\e~ {\rm terms}\right).
\label{eq:a8}
\ee
Note that the first step of the evolution, corresponding to the first term in brackets, has the same form in (\ref{eq:a7}) and (\ref{eq:a8}). However the second step is described by different forms, arising from the different ``+'' prescriptions relevant to diagrams (a) and (b) of Fig. \ref{fig:QEDfig1}. We will give the precise expressions for the $\e$ terms in a moment, but first we discuss the origin of the ``+'' prescription.

\subsection{The + prescription}
The ``+'' prescription is a beautiful way to account for the cancellation of soft photon contributions arising from real emission and from the virtual loop diagrams. In an axial gauge (which is usually used in calculations since it provides a partonic interpretation of the results) the soft photon $1/(1-z)$ singularity arises from the second term in the spin part of photon propagator
\be
d_{\mu\nu}(l)=-g_{\mu\nu}+\frac{l_\mu  n_\nu+l_\nu  n_\mu}{(n \cdot l)}\ ,
\label{eq:spin}
\ee
where the light-cone vector $n=n^-$ is directed along the incoming heavy photon momentum, while $l$ is the momentum of emitted photon which carries a fraction $(1-z)$ of the incoming
electron  momentum $p\simeq p^+$. The $1/(1-z)$ singularity comes from
the factor $(n \cdot l)$ in the denominator of (\ref{eq:spin}).\footnote{Therefore in (\ref{eq:a8}) we get $1/(x_1-x)_+$ and not $1/(1-z)_+=1/(1-x/x_1)_+$ as in (\ref{eq:a7}).} Following \cite{CFP} the singularity is written in the form
\be
\frac 1{l \cdot n}=\frac{l \cdot n}{(l \cdot n)^2+\delta^2(p \cdot n)^2}
\ee
with some very small cutoff $\delta\to 0$. Now if we integrate this singularity over $z$, with any smooth function, then we can use the equality
\be
\frac 1{1-z}=I_0\delta(1-z)+\frac 1{(1-z)_+}
\label{eq:i0}
\ee  
in which
\be
I_0=\int^1_0du\frac u{u^2+\delta^2}\, .
\ee
In this form the singular contribution is collected in a universal function $I_0$, while by introducing the ``+'' prescription in the second (now non-singular) term we account for the precise behaviour at $z\neq 1$. To be sure that all the soft photon singularities are cancelled, we have to check the cancellation of $I_0$ contributions in the final result.  Again following Ref. \cite{CFP}, we may write
\be
\frac {\ln(1-z)}{1-z}=I_1\delta(1-z)+\left[\frac {\ln(1-z)}{(1-z)}\right]_+ ~~~~~~{\rm where}~~~~~I_1=\int^1_0du\frac {u\ln u}{u^2+\delta^2}.
\label{eq:i1}
\ee

\subsection{Difference $\Delta P$ caused by the soft 1/(1-$z$) singularity}
Definitions (\ref{eq:spin}) $-$ (\ref{eq:i1}) were used consistently when calculating the loop contributions in $D=4+\epsilon$ space. Therefore we must use the same definition
(and correspondingly the same form of ``+'' prescription) for the ladder diagram (a) of Fig.\ref{fig:QEDfig1}. Moreover, the same prescription has to be used in diagram (b), so as to provide an exact cancellation of the double log contribution -- both the integrals over $k^2_1$ and $k^2$ gives rise to logarithms. Recall that in the `physical' approach the functions $A$ 
and $A'$ are exactly equal, and so the LO$\otimes$LO subtraction in (\ref{eq:a1}) kills the double log completely. Thus the singularity corresponding to the emission of the upper soft photon is cancelled correctly between diagrams (a) and (c) of Fig.\ref{fig:QEDfig1}.

On the other hand, the above form of the ``+'' prescription, besides being different from that used in the original LO$\otimes$LO calculation, violates the $x/x_1 \Leftrightarrow x_1$ symmetry. So, when the operator $P_{\e n}$ doubles the contribution of the upper cell in Fig.\ref{fig:QEDfig1}(b), but kills the
$\delta_P$ contribution from the lower photon, there is {\it not} an exact compensation in (\ref{eq:a4}). That is, with dimensional regularization there remains a contribution $\Delta P\neq 0$.
In other words, the conventional approach to the NLO splitting function contains some non-physical
 $\e/\e^2$ contribution, $\Delta P$, of infrared 
 origin, which cannot be treated as simply using an alternative factorization scheme, see the discussion in Section \ref{sec:alt}.

Using the notation of (\ref{eq:a5}), this $\e/\e^2$ contribution, $\Delta P$, can be calculated as the difference between the $P^{\rm LO}(x_1)\otimes\delta_P(x/x_1)$ convolutions obtained with the different $1/(1-x/x_1)_+$ and $x_1/(x_1-x)_+$  prescriptions of (\ref{eq:a7}) and (\ref{eq:a8}) respectively.  To be more specific, it means that the $P^{\rm LO}(x/x_1)$ and $\delta_P(x/x_1)$ contributions in (\ref{eq:a5}), will be regularised differently (which below we denote by a prime\footnote{By prime we mean the regularisation $\int \frac{dx_1}{(x_1-x)_+}$ used in the NLO calculation, while at LO without the prime we use $\int \frac{d (x/x_1)}{(x/x_1)}\frac{1}{(1-x/x_1)_+}$. })
to the $\delta_P(x_1)$ and $P^{\rm LO}(x_1)$ contributions. It is convenient to divide $\Delta P$ of (\ref{eq:a5}) into two parts, $\Delta P=\Delta P_1+\Delta P_2$, where 
\begin{equation}
\Delta P_1~=~\delta_P (x_1)\otimes P^{\rm LO'}(x/x_1)-P^{\rm LO}(x_1)\otimes \delta_P(x/x_1),
\label{eq:p1}
\end{equation}
\begin{equation}
\Delta P_2~=~P^{\rm LO}(x_1)\otimes \delta_P (x/x_1)-P^{\rm LO}(x_1)\otimes \delta_P^\prime(x/x_1).
\label{eq:p2}
\end{equation}
Note that the $P^{\rm LO}(x_1)\otimes \delta_P(x/x_1)$ term has been subtracted in $\Delta P_1$ and added in $\Delta P_2$.

First, we evaluate $\Delta P_1$ due to the difference of the first set of brackets in (\ref{eq:a7}) and (\ref{eq:a8}), which corresponds to the second step in the evolution,
\begin{eqnarray}
\Delta P_1&=&\int dx_1 \frac{f(x_1,x)}{(x_1-x)_+}- \int \frac{d (x/x_1)}{(x/x_1)}~ \frac{f(x_1,x)}{(1-x/x_1)_+} \label{eq:b2}\\ 
 &=&\int dx_1 \frac{f(x_1,x)-f(x,x)}{(x_1-x)}- \int \frac{d x_1}{x_1} \frac{f(x_1,x)-(x/x_1)f(x,x)}{(1-x/x_1)} \\
&=&\ - \int_x^1\frac{dx_1}{x_1}f(x,x) = \ln x f(x,x),
\end{eqnarray}
where
\be
f(x_1,x) = (1+(x/x_1)^2)~[(1-x_1)+P^{\rm LO} (x_1)\ln(1-x_1)],
\ee
see $\delta_P (x_1)$
 of (\ref{eq:3terms}). At the moment, we are including only real two photon emission, so the final term in (\ref{eq:3terms}) is omitted. The self-energy terms are included in (\ref{eq:2v}).  Note that the variable of integration in the second term of (\ref{eq:b2}) is written in the form necessary for the application of $+$ prescription. 
 
For the second contribution, $\Delta P_2$, the first term of $\delta_P$ of (\ref{eq:3terms}) does not contribute, since it does not contain a divergence in $x/x_1 \rightarrow 1$, therefore it is the same in $\delta_P'(x/x_1)$ and $\delta_P(x/x_1)$ and cancels out. So we need to evaluate 
\begin{equation} 
\Delta P_2 =  \int_x^1 \frac{d (x/x_1)}{(x/x_1)}  g(x_1,x) \frac{\ln (1 - x/x_1)}{(1-x/x_1)} 
- \int_x^1 d x_1 g(x_1,x) \frac{\ln (x_1-x)}{(x_1-x)}
+ \int_x^1 d x_1 g(x_1,x) \frac{\ln x_1}{(x_1-x)},
\label{eq:pp2}
\end{equation}
where $g(x_1,x)=(1+(x/x_1)^2)P^{\rm LO}(x_1)$. The presence of the logarithms in the numerators of (\ref{eq:pp2}) makes the evaluation more subtle. Eq.(\ref{eq:pp2})  is the unregularised form. We use the + prescription to regularise each of the above three integrals, and obtain the three expressions given below in large (...) brackets, respectively.
\begin{align} \nonumber
 \Delta P_2 = &\left( \int_x^1 \frac{d (x/x_1)}{(x/x_1)}  g(x_1,x) \left[ \frac{\ln (1 - x/x_1)}{(1-x/x_1)} \right]_+ - \int_{x/x_1=0}^{x/x_1=x} d (x/x_1) g(x,x) \frac{\ln (1-x/x_1)}{(1-x/x_1)} + I_1 g(x,x)\right)\\ \nonumber
 & - \left(\int_x^1 d x_1 g(x_1,x) \left[ \frac{\ln (x_1-x)}{(x_1-x)}\right]_+  - \int_1^{1+x} d x_1 g(x,x) \frac{\ln (x_1-x)}{(x_1-x)} + I_1 g(x,x) \right) \\
 & + \left( \int_x^1 d x_1 g(x_1,x) \frac{\ln x_1}{(x_1-x)_+} - \int_1^{1+x} d x_1 g(x,x) \frac{\ln x}{(x_1-x)} + I_0 \ln x g(x,x)\right).
 \label{eq:reg3} 
\end{align}
Note that the $x_1$ interval $(1,1+x)$ has been subtracted, since the $u$ interval is $(0,1)$ in (\ref{eq:i0}) and (\ref{eq:i1}). Analogously the interval of $x_1$ from 0 to $x$ has been subtracted when the ratio $x/x_1$ plays the role of the $u$-variable. From (\ref{eq:reg3}) we have
\begin{align} \nonumber
 \Delta P_2 =& - \int_x^1 d (x/x_1) g(x,x) \frac{\ln (1-x/x_1)}{(1-x/x_1)} 
 + \int_x^1 d x_1 g(x,x) \frac{\ln (x_1-x)}{(x_1-x)} \\
 &  - \int_x^{1+x} d x_1 g(x,x) \frac{\ln x}{(x_1-x)} + I_0 \ln x  g(x,x).
\end{align}
Finally, performing the integration over $x_1$ we obtain 
\begin{align} \nonumber
\Delta P_2=\left[ \frac{1}{2} \ln^2 x + I_0 \ln x \right] g(x,x),
\end{align}
 and hence  for real two-photon emission
 \be
 \Delta{\tilde P}_{\rm real}=\Delta P_1+\Delta P_2= 2 (1-x) \ln x + 2
\frac{1+x^2}{1-x} \ln (1-x) \ln x + \frac{1+x^2}{1-x} \ln^2 x + 2
\frac{1+x^2}{1-x} \ln x I_0.
\label{eq:DP}
\ee
The function $\Delta{\tilde P}_{\rm real}$ exactly coincides with the difference between the real two-photon emission calculated conventionally, with the LO$\otimes$LO term subtracted (4th column of Table 1 in \cite{CFP}), and that calculated in $D=4$ dimensions using the `physical' approach.

Recall that this is the difference for real two-photon emission, whereas in the LO$\otimes$LO subtraction we need to account for the whole LO contribution, which can be written as the real emission regularised by the + prescription, see (\ref{eq:i0}) and (\ref{eq:i1}). In other words, we have to omit the $I_0$ contribution in (\ref{eq:DP}), which is actually cancelled by the LO self-energy terms. So finally we  get
\be
 \Delta P_{\rm real}=\Delta P_1+\Delta P_2= 2 (1-x) \ln x + 2
\frac{1+x^2}{1-x} \ln (1-x) \ln x + \frac{1+x^2}{1-x} \ln^2 x.
\label{eq:DP2}
\ee

Of course, an analogous problem occurs in the NLO self-energy terms. Here the $\epsilon/\epsilon$  correction may be calculated simply, just using the (particle) conservation laws. We obtain
\begin{equation}
\Delta P_{\rm virtual}(x) = \delta(1-x)\int dx'~\Delta P_{\rm real}(x').
\label{eq:2v}
\end{equation}

\section{The numerical size of the discrepancy \label{sec:5}}

To demonstrate the size of the correction numerically, we calculate the $\gamma^*e \to eX$ cross section starting from the simple input $e(x)=x (1-x)$ in the `physical' scheme, and evolve it over particular $Q^2$ intervals; we recalculate the corresponding input $e'(x)$ in the conventional $\overline{\rm MS}$ scheme (see (\ref{eq:eprime})) so as to have the same function $F_2$ at the starting scale $Q^2=Q_0^2$.  Bearing in mind the interest in an analogous QCD effect, we replace the coupling $\alpha_{\rm QED}$ by $\alpha_s(Q^2)$, and consider evolution from $Q^2=Q^2_0=4~{\rm GeV}^2$ up to $Q^2=20$ and $Q^2=500~ {\rm GeV}^2$. 

\vspace{-0.0cm}
\begin{figure} [htb]
\begin{center}
\includegraphics[height=10cm]{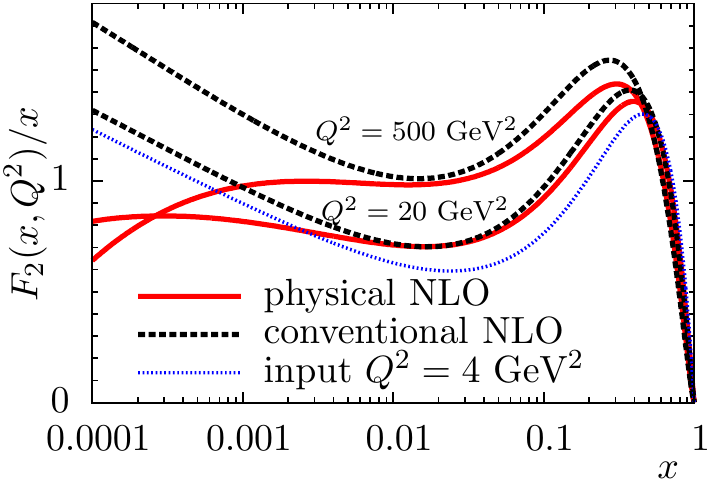}
\vspace*{-0.5cm}  
\caption{\sf The difference between the non-singlet structure function for $\gamma^* e \to eX$ for two values of $Q^2$, calculated using the `physical' and `conventional' treatments of the NLO correction. We use the QCD coupling $\alpha=\alpha_s(Q^2)$, in order to indicate potential effects in the relevant  application to QCD. We evolve up from $Q^2=4~{\rm GeV}^2$ starting from the same $F_2$ in both treatments (shown by the dotted curve). To make the low $x$ behaviour more visible we plot $F_2/x$, since the non-singlet function $F_2(x) \to 0$ as $x \to 0$. }
\label{fig:F2}
\end{center}
\end{figure} 

The results are shown in Fig. \ref{fig:F2}. They are presented in the traditional form of the non-singlet structure function $F_2(x,Q^2)$ as a function of $x$, for the two values of $Q^2$. The solid and dashed curves correspond to the NLO physical and conventional treatments of the infrared region respectively. The difference is clearly evident at low $x$, and increases with $Q^2$. 

Of course, for QED, with its very small coupling $\alpha_{\rm QED} \simeq 1/137$, the effect is tiny. However the estimates shown are relevant to indicate possible effects for QCD. 

{\bf To conclude,} the QED calculation was done here in order to simplify the discussion and to demonstrate that there is a true `physical' correction which cannot be hidden by the re-definition arising from a particular choice of factorisation scheme. The corrections to the coefficient and splitting functions do not compensate each other. Moreover, the correction due to the splitting function depends on $Q^2$ and increases with $Q^2$, especially at very low $x$.  We emphasize that a correction of ${\cal O}(10\%)$  to the conventional treatment of the infrared region was already found \cite {OMR} in the QCD NLO $\gamma^* g$ coefficient in DIS (and in the NLO $qg$ coefficient function for Drell-Yan processes)  using the `physical' treatment of the infrared region.

\section*{Acknowledgements}

EGdO and MGR thank the IPPP at the University of Durham for hospitality. This work was supported by the grant RFBR 11-02-00120-a
and by the Federal Program of the Russian State RSGSS-4801.2012.2;
and by FAPESP (Brazil) under contract 2012/05469-4.

\vspace*{1cm}

\fbox{%
\begin{minipage}{14cm}

{\bf Erratum:} The final conclusion of this paper is not correct.  It was based on our
 misunderstanding of the treatment of the subtraction term to the NLO
 ladder diagram given in column 4 of table 1 of the original calculation of Curci, Furmanski, Petronzio  \cite{CFP}.  Contrary to our treatment, Curci et al. do not use the `+'
 prescription to calculate the NLO contribution in $4+2\epsilon$
 dimensions. Instead they explicitly trace the cancellation of the
 singularities caused by the soft gluon emissions between the real and
 virtual contributions. Thus, it turns out, that the non-singlet
 structure function and the DGLAP evolution in the physical scheme
  are exactly the same as in the ${\overline {\rm MS}}$ scheme. Therefore the physical
 approach may be considered as an alternative factorisation scheme.
\end{minipage}}


\thebibliography{}

\bibitem{BvH} B. Humpert and W.L. van Neerven, Phys. Lett. {\bf B84}, 327 (1979).

\bibitem{BvH2} B. Humpert and W.L. van Neerven, Phys. Rev. {\bf D24}, 2245 (1981).

\bibitem{BN} F.~Bloch and A.~Nordsieck, Phys.\ Rev.\ {\bf 52}, 54 (1937).

\bibitem{LN} F.E. Low, Phys. Rev. {\bf D12}, 163 (1975);\\
S. Nussinov, Phys. Rev. Lett. {\bf 34}, 1286 (1976).

\bibitem{BFKL} 
  L.~N.~Lipatov,  Sov.\ J.\ Nucl.\ Phys.\  {\bf 23} 338 (1976)
   [Yad.\ Fiz.\  {\bf 23} 642 (1976)];\\
  V.~S.~Fadin, E.~A.~Kuraev and L.~N.~Lipatov,
  Phys.\ Lett.\ B {\bf 60} 50 (1975);\\
  E.~A.~Kuraev, L.~N.~Lipatov and V.~S.~Fadin,
  Sov.\ Phys.\ JETP {\bf 44} 443 (1976)
   [Zh.\ Eksp.\ Teor.\ Fiz.\  {\bf 71} 840 (1976)];\\
  E.~A.~Kuraev, L.~N.~Lipatov and V.~S.~Fadin,
  Sov.\ Phys.\ JETP {\bf 45} 199 (1977)
   [Zh.\ Eksp.\ Teor.\ Fiz.\  {\bf 72} 377 (1977)].
  
\bibitem{OMR} E.G. de Oliveira, A.D. Martin and M.G. Ryskin, JHEP {\bf 1302}, 060 (2013).

\bibitem{CFP} G. Curci, W. Furmanski and R. Petronzio, Nucl. Phys. {\bf B175}, 27 (1980).

\bibitem{AEM} G. Altarelli, R.K.Ellis and G. Martinelli, Nucl. Phys. {\bf B157}, 461 (1979).

\bibitem{FORM} J.~A.~M.~Vermaseren, math-ph/0010025.
\end{document}